\documentclass[12pt]{iopart} 

\usepackage{epsfig} 
 
\def\eg{{e.g.},~}   
\def\etal{{et al.}~}

\newcommand{\rsn}{{\rm R}_{\rm SN}} 
\newcommand{\msun}{{\rm M}_\odot} 
\newcommand{\yr}{{\rm yr}} 
\newcommand{\mpc}{{\rm Mpc}} 
\newcommand{\sfr}{\dot \rho_*} 
\def\lsim{\lower0.6ex\vbox{\hbox{$ \buildrel{\textstyle <}\over{\sim}\ $}}}   
\def\gsim{\lower0.6ex\vbox{\hbox{$ \buildrel{\textstyle >}\over{\sim}\ $}}}   
\def\SK{Super-K} 
\def\kam{KamLAND}  
\def\snii{SN\thinspace{$\scriptstyle{\rm II}$}}

\def\epm{$e^{\pm}\;$}  
\def\barnue{\bar{\nu}_{\rm e}}   
\def\barnumu{\bar{\nu}_{\mu}}   
\def\barnutau{\bar{\nu}_{\tau}}   
   
\def\numu{\nu_{\mu}}   
\def\nutau{\nu_{\tau}}   
\def\eplus{{\rm e}^{+}}   
   
\def\Fbare{{\rm F}_{\bar{{\rm e}}}}   
\def\Fbare0{{\rm F}_{\bar{{\rm e}}}^{{\rm o}}}    
\def\Fbarx0{{\rm F}_{\bar{{\rm x}}}^{{\rm o}}}    
\def\Ue1{\vert {\rm U}_{{\rm e}1} \vert}   
\def\Ue3{\vert {\rm U}_{{\rm e}3} \vert}   
\def\Tnubare{\langle {\rm T}_{\bar{\nu}_{{\rm e}}} \rangle}   
   
\def\Tnubarmu{\langle {\rm T}_{\bar{\nu}_{\mu}} \rangle}   
\def\Tnubartau{\langle {\rm T}_{\bar{\nu}_{\tau}} \rangle}

\begin{document}   
   
\title{The Supernova Relic Neutrino Backgrounds at KamLAND and 
  Super-Kamiokande}     
   
\author{Louis E. Strigari$^{1}$, Manoj Kaplinghat$^{3}$, Gary  
Steigman$^{1,2}$, and Terry P. Walker$^{1,2}$}   
 
 \address{$^1$Department of Physics, The Ohio State University,    
Columbus, OH 43210}   
\address{$^2$Department of Astronomy, The Ohio State University,   
   Columbus, OH 43210}   
\address{$^3$Department of Physics, University of California, 
Davis, CA 95616}   
   
\ead{strigari@mps.ohio-state.edu, kaplinghat@physics.ucdavis.edu,  
steigman@mps.ohio-state.edu, and twalker@mps.ohio-state.edu}

\date{\today} 
   
\begin{abstract}    
We calculate  the Supernova Relic Neutrino (SRN) background  flux for 
the  KamLAND and  Super-Kamiokande (\SK)  detectors, motivated  by the 
reduction in background at \SK~and new results for the star formation 
history (e.g.,  from the Sloan  Digital Sky Survey (SDSS)).   Our best 
estimate for the flux at \SK~is  slightly below, but very close to the 
current \SK~upper limit.  The  \SK~upper limit is already inconsistent 
with a range of star formation  histories allowed by the SDSS data. We 
estimate that the SRN background should be detected (at 1$\sigma$) at 
\SK~with a total  of about 9 years (including the  existing 4 years) 
of data.  While  KamLAND is a much  smaller detector compared  to \SK, 
it profits   from  being   practically  background-free   and   from 
its sensitivity to  the lower  energy supernova neutrinos.   KamLAND 
could make a 1$\sigma$  detection of the SRN with a total  of about 5 
years of data.  Given the small expected SRN event rate, we also 
consider  the detection of the  SRN in a modified \SK~detector with a 
lower threshold and  reduced background where the time to detection 
can be reduced by a factor of 10 relative to the existing \SK~estimate. 
 
\end{abstract}   
   
\maketitle 
 
\section{Introduction} 
In its  death throes  as a  type II supernova  (\snii) a  massive star 
($\gsim 8  {\rm M}_\odot$) ends its  life emitting $\sim  99\%$ of its 
energy,   $\sim  10^{53}$~ergs,   in   neutrinos.   This   theoretical 
expectation was spectacularly confirmed  by the detection of a handful 
of   neutrinos   from  SN1987A   in   the   nearby  Large   Magellanic 
Cloud~\cite{hirata,bionata}.   While  the  much larger,  more  heavily 
instrumented detectors such as  Super-Kamiokande (\SK) and \kam~ await 
the  flood  of  neutrinos  from  another  such  nearby  \snii,  it  is 
interesting  to ask  if  either detector  might  observe the  diffuse, 
isotropic  flux of  neutrinos from  all \snii  ~events that  have ever 
occurred within the observable universe. 
 
The  detection  of   this  cosmological  background of core-collapse 
supernova relic neutrinos (SRN) offers a new probe of \snii~neutrino 
physics and of the high  redshift  Universe.  The basic picture of 
core-collapse \snii~could be tested, not only locally but also at high 
redshifts ($z ~\gsim 1$).  From  the SRN we will also obtain estimates 
of the  supernova rate (proportional to  the star formation rate for  
$M~\gsim 8  {\rm M}_\odot$) and of  the metal enrichment rate which  
are completely  independent of those from optical and UV surveys.   
Note that  other  tracers of the cosmological  star  formation  history 
typically  only lead  to  lower  bounds on  the  star formation  rate. 
Comparison with traditional methods would yield information about high 
redshift  star formation  complementary  to  that  from  future  high 
redshift galaxy surveys like  DEEP2 \cite{davis02} (which  targets $z 
~\lsim 1$).  Here we consider the possibility of SRN detection at two 
existing  neutrino detectors, \SK~and  \kam.   We also consider the 
detection of  the SRN  in a modified \SK~detector with lower threshold   
and reduced  backgrounds  (like the  recently  proposed SK-GADZOOKS  
\cite{GADZOOKS}).  As discussed later, detection  of the SRN at {\it both}  
\kam~and \SK~can  be used to probe the cosmic star formation  history  
at $z ~\gsim 1$, of  which little is known from traditional astronomical  
methods. 
 
The prediction of the SRN flux has been the subject of many previous 
investigations \cite{previous}.  The earlier study of Kaplinghat,  
Steigman, and Walker 2000 (KSW) \cite{ksw} reached the pessimistic  
conclusion that it would be unlikely for \SK~to detect the SRN background,  
a result driven by three factors.  First there was the realization that  
for \SK~to detect these relic neutrinos, the actual flux must be close  
to the predicted upper bound and, in addition, the $\barnue$ must be  
nearly maximally mixed with $\barnumu$ or $\barnutau$.  However, as  
discussed below, recent results on the cosmic star formation history  
from observational data, including those from the Sloan Digital Sky  
Survey (SDSS), lead to a predicted SRN rate that {\it is} closer to  
the {\it upper} bound derived in KSW.   Furthermore, neutrino experiments,  
such as SNO, \kam~, and \SK~have clearly identified the large mixing  
angle case  as the  preferred solution  to the  solar neutrino problem  
\cite{kam2, sno}, implying  that the \snii~$\bar{\nu}_{e}$ {\it are} 
(nearly) maximally mixed.  The second factor concerned the backgrounds 
to the detection  of the SRN. KSW estimated  the primary background at 
\SK,  due   to  sub-Cherenkov  muons  (to  be   discussed  later),  by 
extrapolating from older Kamiokande data.  The estimated background to 
the detection of the SRN in the energy window (for  positrons) from 19 to 
35  MeV was  39 events  per year.   The latest  \SK~analysis \cite{SK} 
reveals  that the  background in  the same  energy window  is actually 
about 19  events per year.  The  third factor leading  to the negative 
conclusion of KSW concerned the  spectrum of the SRN flux.  The difference 
between the  spectra of  the SRN and  the sub-Cherenkov  muon neutrino 
fluxes could help to separate the SRN signal from this background but, 
given  the large  expected  background and  the  uncertainties in  the 
supernova  rate, KSW  did not  attempt to  account for  this  in their 
analysis.  However,  considering the current  status of the  first two 
factors, it is clear that a more detailed analysis is now appropriate. 
Although  with 4 years  of data  \SK~did not  see any  supernova relic 
neutrinos, they  did reduce the  upper limit on  the SRN flux  at 90\% 
C.L. to $1.2 ~\barnue$  cm$^{-2}$s$^{-1}$, for positron energies $> 18  
\hspace{.1cm}  {\rm  MeV}$~\cite{SK}.  This  limit  is  more than  two 
orders  of magnitude  smaller  than the  previous Kamiokande-II  upper 
limit   \cite{kamiokandeII}  and   is  approaching   several   of  the 
theoretical estimates  of the SRN  flux.  Given this  very encouraging 
state  of affairs,  along with  the advances  in recent  years  in our 
understanding of the  cosmic star formation history, we  believe it is 
timely  to attempt a  realistic, quantitative  estimate of  the likely 
range for the SRN flux. 
 
For \SK~our  analysis is restricted to positron energies $> 18$ MeV. 
In  contrast the KamLAND detector, as a result of their detection 
technique, have much smaller backgrounds at energies suitable for 
detecting SRN and are therefore sensitive to much lower energy 
neutrinos (positrons with E $>  6$ MeV).  Thus, even though KamLAND  
is a much smaller detector than \SK~, it profits with respect to  
\SK~by being sensitive to more of the SRN spectrum and also by its  
sensitivity to neutrinos from supernovae at higher redshifts compared  
to \SK.  We also present results for the SRN flux expected at KamLAND  
in the idealized limit in which KamLAND is able to utilizes its entire  
1 kton volume for detecting SRN with 100\% efficiency. 
We also discuss detection in the hypothetical SK-GADZOOKS detector, the  
propeties of which were described in Beacom \& Vagins \cite{GADZOOKS}. 
Estimates for a hypothetical detector ``HyperKamiokande'' \cite{hyperK},   
with a fiducial volume of $V  \approx 890$  kton($\approx 40V_{\rm SK}$),  
can be obtained by appropriately scaling the flux at \SK~and noting that  
the errors on the flux scale as $V^{-1/2}$.     
 
\section{Detectors} 
\SK~is a 22.5 kton fiducial mass water Cerenkov detector which can detect 
SRNs  via inverse  beta decay  $\barnue  + {\rm  p} \rightarrow  {\rm 
e}^{+}  + {\rm  n}$.   Below  $18 \, {\rm  MeV}$ spallation  events 
resulting from the interaction of cosmic ray muons with oxygen nuclei  
form the primary background at \SK~\cite{Gando}.  Above $18 \, {\rm MeV}$   
there are two primary backgrounds.  The first comes from low energy  
atmospheric $\barnue$ produced from the decay chain of $\pi^{\pm}$ and  
$\mu^{\pm}$  \cite{Atmospheric}.  Below  $25  \, {\rm MeV}$ the SRN  
flux from our median estimate (see below), which peaks at roughly $5  
\, {\rm MeV}$, exceeds the atmospheric $\barnue$ flux.  The second  
background comes from atmospheric $\nu_{\mu}$ which interact with a  
nucleus to form a $\mu^{\pm}$ with kinetic energy $\le 53 \, {\rm MeV}$.  
Such $\mu$s are not detected as they are below the threshold for  
Cerenkov radiation, but their \epm decay products are detected, with  
energy distribution given by the Michel spectrum.  By performing a 
multicomponent fit to the observed Michel spectrum, as well as that  
for the atmospheric $\nu$, \SK~obtained their impressive limit on the  
SRN background  flux \cite{SK}.  It would be ideal if there were a  
way to detect the recoil neutrons from the  $\barnue - p$ interaction  
(analgous to  the  neutral  current detection at SNO) as was suggested  
in a recent proposal \cite{GADZOOKS}.    
This  neutron tagging  would  remove  the $\nu_{\mu}$ and spallation 
backgrounds, allowing \SK~to move to a lower energy threshold of 10 MeV.  
Such a reduction in threshold will tell us  more about high  redshift  
supernovae  and the corresponding star  formation history than any   
other existing method. 
Regardless, for the estimates of the SRN flux at \SK~presented here we 
limit ourselves to neutrinos with  energies $ > 19.3$~MeV (positrons 
with  E$_{e^{+}} >  18$~MeV).  The current predictions from Ando, Sato,  
\& Totani \cite{previous} show that for SK the expected flux is roughly  
$0.3 \, {\rm cm^{-2}} {\rm s^{-1}}$, which is close to the lower limit  
derived from Fukugita \& Kawasaki \cite{previous}.  
Below we show how median estimates  
of the SN rate as a function of redshift can give a flux very close to  
the current \SK~upper limit.  The corresponding event rate of the SRN  
at \SK~in this range is expected to be 1 to 2 per year, and a few more  
years of data are likely to yield a positive detection.  
 
KamLAND is a  1 kton liquid scintillation detector  designed to search 
for  evidence  of  $\barnue$  oscillations  utilizing  $\barnue$  from 
nuclear   power   reactors   \cite{kam1,bem,kam2}.   KamLAND   detects 
$\barnue$  via inverse  beta decay  by a  prompt signal  from positron 
annihilation, followed  by a  $\sim 200 \,  \mu {\rm  s}$ time-delayed 
neutron capture $\gamma$-ray of 2.2  MeV.  The spectrum of the reactor 
$\barnue$  peaks  near  3  MeV,  and  with  neutrino  oscillations  is 
negligible above 6 MeV.   The KamLAND collaboration analysis imposes a 
lower cutoff of 2.6 MeV to account for the background from terrestrial 
$\barnue$ sources.  Unlike \SK, KamLAND can readily exclude the invisible 
muon decay background since the prompt signal from positron annihilation 
is not followed by neutron capture.   The KamLAND background at energies 
greater than 6 MeV is from atmospheric $\barnue$, which competes with 
our estimated SRN signal at energies $\gsim \, 25 {\rm} \,  {\rm MeV}$. 
KamLAND's smaller background at low energies, in the region where the 
SRN spectrum peaks, may make it possible for KamLAND to detect the SRN 
background.  Preliminary results at \kam~ for a 0.28 kton-year exposure 
show no $\barnue$ signal above the small expected backgrounds \cite{KamBack}.  
Current theoretical estimates assuming that KamLAND can use the entire 1 
kton volume for SRN detection (Ando, Sato and Totani 2003~\cite{previous})  
suggest that in the energy window 10 -- 25 MeV, the SRN event rate is 
$\sim 0.1~\barnue~{\rm  yr}^{-1}$.  In our analysis here the SRN flux 
for (positron) energies E $> 6$ MeV is considered.  This lower energy  
threshold is chosen since, so far, there are  no events seen above this 
energy at KamLAND~\cite{kam2}.  In addition, in our estimates for KamLAND 
we adopt an idealization in which KamLAND utilizes the entire 1 kton 
fiducial volume with $100\%$ efficiency for SRN. Based on our best estimate 
of the supernova rate which saturates the \SK~bound, the expected rate in an 
idealized KamLAND detector is 0.4 events per year.
 
\section{Flux of the SRN} 
The  differential  flux of  SRN  $\barnue$,  $dF/dE$,  depends on  the 
magnitude  and  evolution (as  a  function  of  redshift $z$)  of  the 
\snii~supernova rate $R_{SN}(z)$, and on the $\barnue$ energy spectrum 
$dN/dE$, 
 
\begin{equation}  
\frac{dF}{dE}        =        \int_{0}^{z_{max}}\rsn(z)        \langle 
\frac{dN(E(1+z))}{dE} \rangle (1+z)| \frac{dt}{dz} | dz. 
\label{flux}  
\end{equation}  
The energy of the detected  positron, $E_{\eplus}$, is related to that 
of the  $\barnue$ by $E_{\eplus} =  E - 1.3  \hspace{.1cm} {\rm MeV}$. 
In the  above equation  the average is  over the stellar  initial mass 
function (IMF $\equiv dn_*/d\ln M$, where $n_*$ is the number of stars 
of mass $M$).   In practice, $\langle dN(E') /  dE\rangle$ is replaced 
by $dN(E') / dE$, with  the parameters that determine the SN $\barnue$ 
spectrum  replaced  by  their  respective IMF-averaged  values.   This 
approximation is accurate to better  than $10 \%$ in the energy window 
($ >  6 \,  {\rm MeV}$) of  interest. To  compute $dt/dz$, we  use the 
$\Lambda$CDM cosmology (i.e.,  a flat, cosmological constant dominated 
model with matter density $\Omega_M=0.3$ and Hubble constant $h=0.7$). 
 
\subsection{Estimating the Supernova Rate} 
 
The  core-collapse  SN rate  at  a given  redshift  is  just the  star 
formation rate (SFR)  at that redshift for stellar  masses larger than 
$8  \, M_{\odot}$.  To  obtain the SN rate we  link the  SFR to  an 
observable proxy such as the UV  or H$\alpha$  luminosity  density.  The 
interpretation of the measured high redshift UV luminosity density is 
complicated by  the fact that UV  light is strongly  absorbed by dust, 
while measurements of H$\alpha$, though less affected by dust, are not 
as  simple  to relate  to  the star  formation  rate.  In addition  to 
correcting  for dust  extinction, converting  the  UV light  to a  SFR 
requires  correcting for  the  incomplete sampling  of the  luminosity 
function,  as well  as  cosmological surface  brightness dimming,  the 
latter  being  true  at   all  frequencies  for  any  extended  object 
\cite{lanzetta02, conti03, rowan-robinson03,thompson03} 
 
>From the determination of the SFR we extract the SN rate, 
$\rsn(z) = \int_{8\msun}^{30 \msun} \sfr(z) dn_*/dM(M') dM'$,  
where $\sfr(z)$ is the star formation rate in average mass per time 
per comoving volume. We have limited the upper bound to 30  
$\msun$ due to the potential  
uncertainties in the neutrino flux from SNe with progenitors more 
massive than that. Here $dn_*/d\ln M$ is the global stellar IMF, assumed 
constant over redshift, which is a good approximation provided  
there are no significant 
correlations between the IMF and the environment in which the stars 
are born. Extant evidence seem to argue against such correlations 
 over the redshift range of interest  ($z \lsim 2$)\cite{scalo98}. Averaging 
over a Salpeter IMF for $M > 8\, \msun$, the SN rate is    
${\rm R}_{{\rm SN}} = (\frac {0.013}{{\rm M}_{\odot}})\sfr$, assuming the star 
formation rate is measured in solar masses.  The   
conversion factor is not sensitive to the upper limit of the average  
as long as the upper limit is larger than about 25 $\msun$. The 
sensitivity to changes in the IMF slope is more pronounced; if we use 
Baldry and Glazebrook \cite{baldry03} (BG03; see below) best-fit IMF 
($dn_*/d\ln M \sim M^{-1.15}$), the conversion  factor increases 
by 30\%.   
 
Following \cite{hogg01,baldry02}, we parametrize the SFR as  
$\sfr(z) \propto (1+z)^{\beta}$ for $z < 1$, and  
$\sfr(z) \propto (1+z)^{\alpha}$ for $z > 1$ and use various 
observational proxies to estimate $\alpha$, $\beta$, and the 
normalization. From our assumption above of a redshift-independent 
IMF, we can then parametrize the SN rate as  
\begin{eqnarray} 
\rsn  & \propto & (1+z)^\beta  \quad {\rm for} \quad z<1 \nonumber \\ 
      & \propto & (1+z)^\alpha \quad {\rm for} \quad z>1 \label{rsn}.  
\label{eq:SNrate}  
\end{eqnarray} 
In Equation \ref{rsn} we have assumed that the $1<z<2$ 
behavior continues to higher redshift.  
 
Hogg \cite{hogg01} has compiled measurements of the UV and H$\alpha$  
luminosity density, as well as results from measurement of the near 
UV emission, far-infrared and radio continuum to obtain the $68 \%$ 
c.l. limits of $\beta = 2.7 \pm 0.7$. Results from the cosmic optical 
spectrum measurements from the Sloan Digital Sky Survey (SDSS) find 
limits on $\beta$ from $2-3$ and $\alpha$ from $0-2$ \cite{glazebrook03} 
(G03).  Using the SDDS data and marginalizing over the IMF and evolution 
of the SFR, BG03 determine the local density of SFR to be in the range  
$\sfr(0) =  (0.5 - 2.9) \times 10^{-2}  \msun \yr^{-1} \mpc^{-3}$.  The 
lower portion of this range for the slopes and normalzation is only 
valid, however, if the slope of the global IMF is allowed to vary near 
the BG03 best fit.  This lower portion is also consistent with 
the results of Madau et al. based on the observed UV luminosity density 
of the whole galaxy population \cite{madau98}. However, the Madau et al. 
results do not take into account the correction for surface brightness 
dimming and provide a smaller average correction for dust extinction 
\cite{thompson03}, and thus permit the possibility that the SN rate may  
{\it decline} for $z \gsim 1$. In addition, studies of extra-galactic 
background light have recently shown that current surface brightness  
corrections suggest that $\sim 50\%$ of the high redshift galaxy 
population is yet unobserved \cite{missinggalaxies}.    
 
The choice of how to handle the dust extinction thus dominates the 
uncertainty budget for the high redshift SN rate.  Although the slope 
of the SFR for $z \gsim 1$ is poorly constrained from the SDSS results, 
examination of high redshift UV data \cite{thompson03} shows that the 
SFR remains constant or even increases beyond $z \simeq 1$.  An increase 
in $\alpha$ would not significantly change the {\it observable} SRN 
flux at Super-K, as the energy window $\ge  18 \, {\rm  MeV}$ is not 
very sensitive to the $z >1$ SN rate.  For example, given the 
parameterization in  Eq.~\ref{rsn}, we find that  the 90  \% C.L. 
\SK~upper bound on the flux of the relic neutrinos of 1.2 
${\rm  cm}^{-2}{\rm s}^{-1}$ implies that, for $\beta > 0.5$, $\rsn(0) 
<  2 \times 10^{-4}\,\yr^{-1} \mpc^{-3}$, independent of $\alpha$. 
However, an  increase in $\alpha$ would increase the observable flux 
at KamLAND  (or any other relatively low  threshold detector such 
as  the proposed SK-GADZOOKS), due to its sensitivity to lower 
energy neutrinos. Below we discuss the extent to which a comparison 
of signals between KamLAND and Super-K can constrain the high redshift 
SFR. 

\begin{figure} 
\begin{center} 
\epsfxsize=3.2in 
\epsfbox{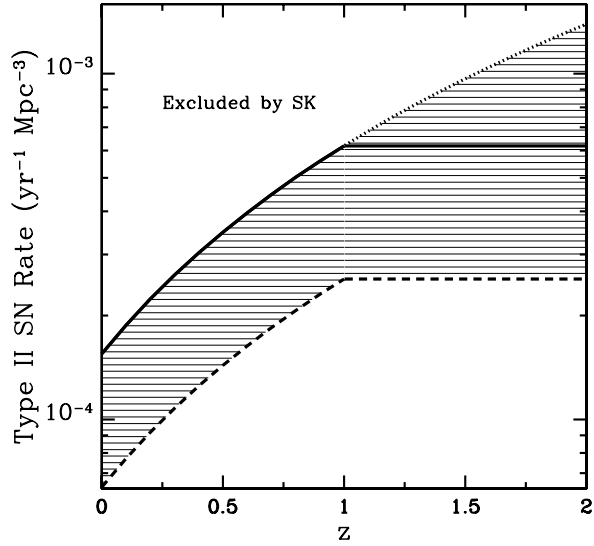} 
\end{center}  
\caption{Three representative supernova rates; see eq.~(\ref{eq:SNrate}). 
The solid curve is based on the median results from UV luminosity density 
measurements as well as the SDSS cosmic optical spectrum.  The dotted
curve is an extension of this median model in which the slope of the 
$z > 1$ SN rate is increased from $\alpha = 0$ to $\alpha = 2$; see
eq.~(\ref{eq:SNrate}).  The area above the shaded region is ruled 
out by the SK flux limit.  The dashed curve is a lower bound to 
the SN rate motivated by the SDSS optical spectrum.  In each case 
a Salpeter IMF is used to convert from the SFR to the SN rate.    
\label{fig:SNrate}} 
\end{figure} 

Motivated by the UV density studies as well as the SDSS cosmic optical 
spectrum, we choose for our ``median/best estimate'' model $\beta = 2.5$, 
$\alpha = 0$, $\sfr(z=0)= 1.6 \times 10^{-2} \, \msun \yr^{-1} \mpc^{-3}$. 
This normalization is perfectly consistent with the observations and 
is specifically chosen to saturate the \SK~result, for the adopted slopes 
$\beta$ and $\alpha$. This ``median'' SN rate (and its associated SRN 
background) is shown by the solid curves in Figures 1 and 2.  Note that 
for this value of $\rsn(z=0)$, and with an average density of galaxies
of $0.01 \, \mpc^{-3}$, the present average supernova rate is expected 
to be $\sim 1$ per 65 years per galaxy.  Though the UV and SDSS analyses 
do permit models with higher star formation rates, such models are ruled 
out by the \SK~upper limits to the SRN flux.  In particular, {\it any 
models} with values of $\beta$ and/or $\rsn(z=0)$ which exceed our 
median estimates will violate the \SK~upper bound on the SRN flux.  
As a specific example, if the slopes were fixed at $\beta = 3$, 
$\alpha = 2$, and a Salpeter IMF adopted, this would require a SFR 
$\sfr(z=0) < 1.3 \times 10^{-2} \, \msun \yr^{-1} \mpc^{-3}$ in order 
to satisfy the \SK~flux limit.  Thus, it is clear that the \SK~upper 
bound on the SRN flux already provides valuable constraints on the 
cosmic star formation history. 
 
It is interesting, but less certain, to estimate a reasonable lower 
limit to the SRN flux based on the above observational proxies.  We 
choose a  ``lower limit'' model, shown by the dotted curves in Figures
1 and 2, with $\beta = 2$, $\alpha = 0$, $\sfr(z=0)= 0.5 \times 10^{-2} 
\, \msun \yr^{-1} \mpc^{-3}$.  Finally, in order to determine the 
impact on the SRN flux of increasing the high redshift SFR, we consider 
a variation to our ``median/best estimate'' model with $\beta = 2.5$ 
and $\alpha =0$, by setting $\alpha = 2$ ( dashed curves in Figures 
1 and 2).  We note that the predictions of Ando, Sato, \& Totani, 
and of Kawasaki \& Fukugita \cite{previous} use SN rates which are 
closer to this SDSS inspired lower bound.    
 
As already mentioned, with its sensitivity to lower energy neutrinos  
KamLAND has the potential to probe the high redshift star formation  
history.  About 40\% of the $\bar{\nu}_{e}$ flux in the KamLAND energy  
window comes from $z>1$; about 10\% of the flux is from  SN at $z>2$.   
Thus it is clear that a comparison of the fluxes at \SK~and KamLAND  
has the potential to provide valuable information about the star  
formation history for $1 < z < 2$. If the SFR at $z ~\gsim 2$ is much  
larger than the estimates adopted here (there are some hints in this  
direction \cite{lanzetta02,conti03}), then it might be possible to probe  
the star formation rate at $z ~\gsim 2$. Below, we  quantify the  
information that can be gleaned about high redshift star formation  
using \SK~and KamLAND.

\subsection{SRN Spectrum} 
The other key ingredient in calculating the background flux of SRN is 
the {\it spectrum} of supernova neutrinos.  The supernova neutrino  
spectrum is typically modelled by a Fermi-Dirac spectrum with an  
effective chemical potential, $\eta \equiv \mu_{\nu}/T_{\nu}$,  
\begin{equation}   
\frac{dN}{d E_\nu} \propto \frac{E_{\nu}^{2}}   
{\exp(E_{\nu}/T_{\nu}-\eta)+1}.    
\label{eq:Fermi}   
\end{equation}    
Equation \ref{eq:Fermi} may be used to relate the average $\barnue$  
energy $\langle E_{\bar \nu_e} \rangle$ to the $\barnue$ neutrinosphere  
temperature.  Our current understanding of the \snii~explosion mechanism  
(and the observations of neutrinos from SN 1987A \cite{hirata, bionata})  
restricts the average energy to the range $14 ~\lsim \, \langle E_{\bar  
\nu_e}\rangle \, ~\lsim \, 17$~MeV (see, \eg \cite{Energies}).  Relative  
to a pure Fermi-Dirac spectrum ($\eta = 0$), for a distribution normalized  
to the total neutrino {\it energy}, positive values for $\eta$ correspond  
to a flattening of the spectrum peak, and an increase in the number of  
neutrinos in the tail of the distribution.  To compromise between the  
extremes seen in computer simulations, we adopt a Fermi-Dirac distribution  
with zero chemical potential, characterized by $\Tnubare = 5 \hspace{.1cm}  
{\rm MeV}$ and $\Tnubarmu = \Tnubartau = 8 \hspace{.1cm} {\rm MeV}$.   
We assume that the total energy carried by each flavor of neutrino is  
$0.5 \times 10^{53} \, {\rm erg}$. The average is taken over the IMF of  
the \snii~progenitors. This does not introduce additional significant 
uncertainties since neither the temperature of the neutrinosphere nor 
the binding energy of the neutron star depend very sensitively on the 
the mass of the progenitor and, hence, on the averaging over the IMF.  
The most recent simulations of Raffelt \emph{et al.} \cite{Raffetal}  
including muon and tau transport models suggest that the IMF-averaged  
energies may not be hierarchical, but that $\langle E_{\bar \nu_e}  
\rangle \, \approx \, \langle E_{\bar \nu_{\mu,\tau}} \rangle$.  For  
a spectrum normalized to the same total energy output in all neutrino 
species, such differences do not significantly change the results for  
the fluxes.  
 
\subsection{Neutrino Mixing}{\label{sec:mixing} 
Neutrino oscillations are of direct relevance to the spectrum of  
background SRN.  If $\barnumu$ and/or $\barnutau$ mix with $\barnue$  
the spectrum of the resulting $\barnue$ will be harder.  Such higher 
energy $\barnue$ are easier to detect.  From the analyses of solar  
neutrino and KamLAND data, mixing between the $\nu_{e} (\barnue)$  
and $\nu_{\mu} (\barnumu)$ flavor eigenstates is favored, with $0.27  
< \tan^{2} \theta_{\odot} < 0.94$ at $3 \sigma$ \cite{Bahcall}.  In  
terms of mass eigenstates, the MSW solution implies $\Delta m_{\odot}^{2}  
\equiv m_{2}^{2} - m_{1}^{2}> 0$ \cite{msw}. The atmospheric neutrino 
data \cite{fuku} are consistent with near-maximal vacuum mixing between 
the $\numu (\barnumu)$ and $\nutau$($\barnutau$) flavor eigenstates: 
$\sin^{2} (2\theta) > 0.82$ at $90\%$ C.L.  However, unlike the solar 
neutrino and \kam~data, the atmospheric neutrino data do not constrain 
the sign of the corresponding mass-splitting, $\Delta m_{atm}^{2} 
\equiv \vert m_{3}^{2}-m_{2}^{2} \vert$.  Here, two mass hierarchies 
are possible; a {\it normal} hierarchy, with $m_{3} > m_{2} > m_{1}$, 
or an {\it inverted} hierarchy, with $m_{2} > m_{1} > m_{3}$.  As will 
be noted next, the choice of hierarchy is important for any neutrino 
mixing inside the supernovae. 
  
Due to the high density inside supernovae, there is the possiblity that  
before reaching the surface of the supernovae, the neutrinos may have  
been mixed by the MSW effect \cite{msw}.  In supernovae, there are two 
MSW resonant density layers, corresponding to the solar and atmospheric 
mass splittings.  Whether the resonant density occurs in the $\nu$ or 
$\bar{\nu}$ channels depends on the mass hierarchy \cite{DS}.  For a 
normal hierarchy, both resonant layers are in the $\nu$ channel.  
However, the flux that arrives at the surface of the supernova is 
an incoherent mixture of mass eigenstates.  These mass eigenstates  
travel separately to the surface of the earth and, in the absence 
of earth matter effects, the final $\barnue$ flux will have a fraction 
$\sin^{2} \theta_{\odot}$ of the original $\bar{\nu}_{\mu}$ flux.} 
In contrast, for an inverted hierarchy, the higher density resonant 
layer is in the $\bar{\nu}$ channel.  Deep inside the supernova the 
$\barnue$ flavor corresponds to the lowest mass eigenstate, whereas 
in vacuum $\barnutau$ is the lowest mass eigenstate.  For adiabatic  
propagation inside the supernova, and for $\Ue3 \sim 0$ (but not  
exactly zero) as implied by the reactor neutrino data \cite{bem},  
$\barnue$ remains the lowest mass eigenstate as it exits the supernova, 
while $\barnutau$ remains the heavy mass eigenstate.  The result is 
that for an inverted hierarchy the $\barnue$ observed on Earth were 
all ``born'' as $\barnutau$.   
 
\section{Results} 
 
\begin{figure}[b] 
\begin{center} 
\hbox{\epsfig{file=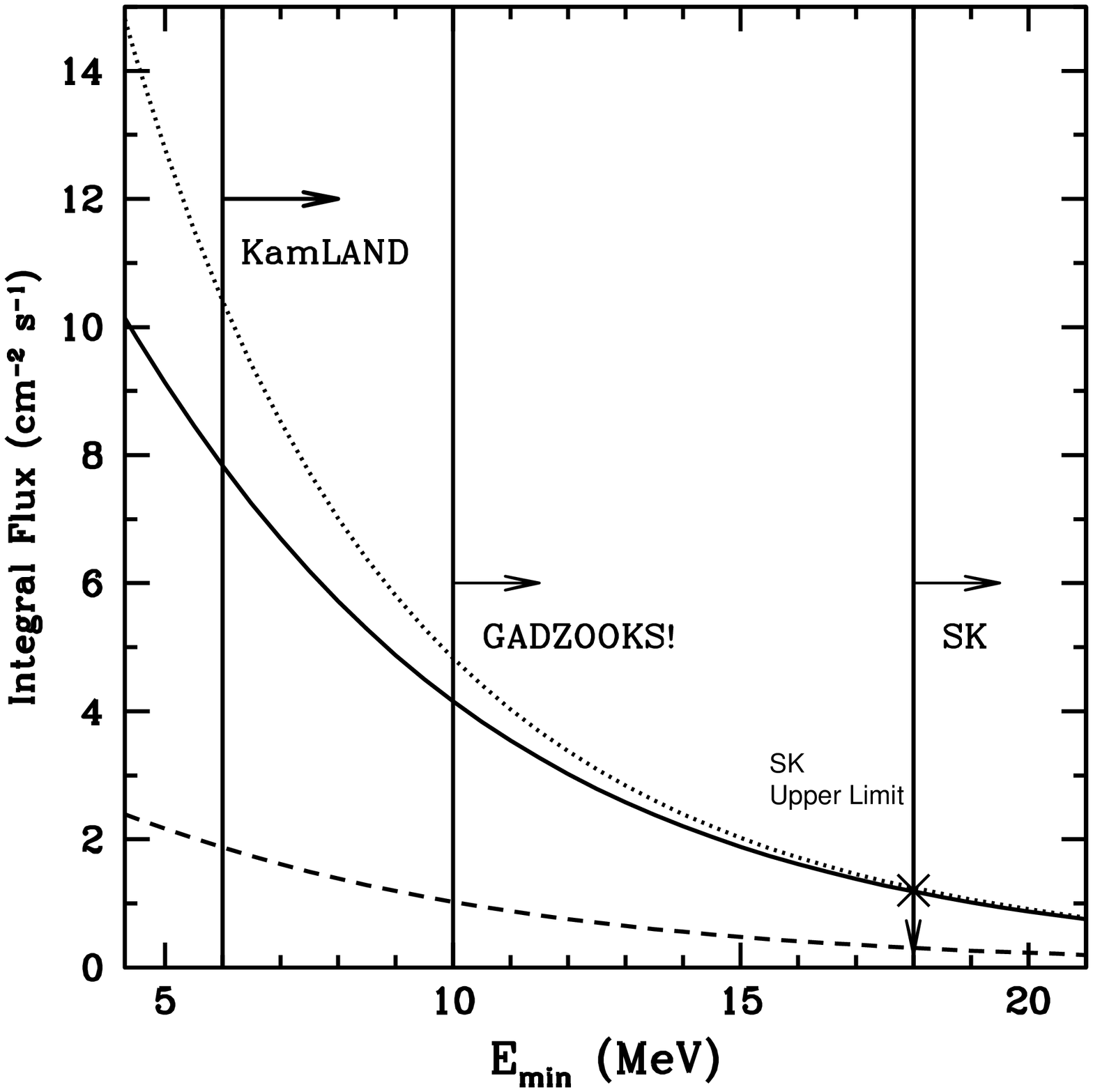,width=8cm} 
\epsfig{file=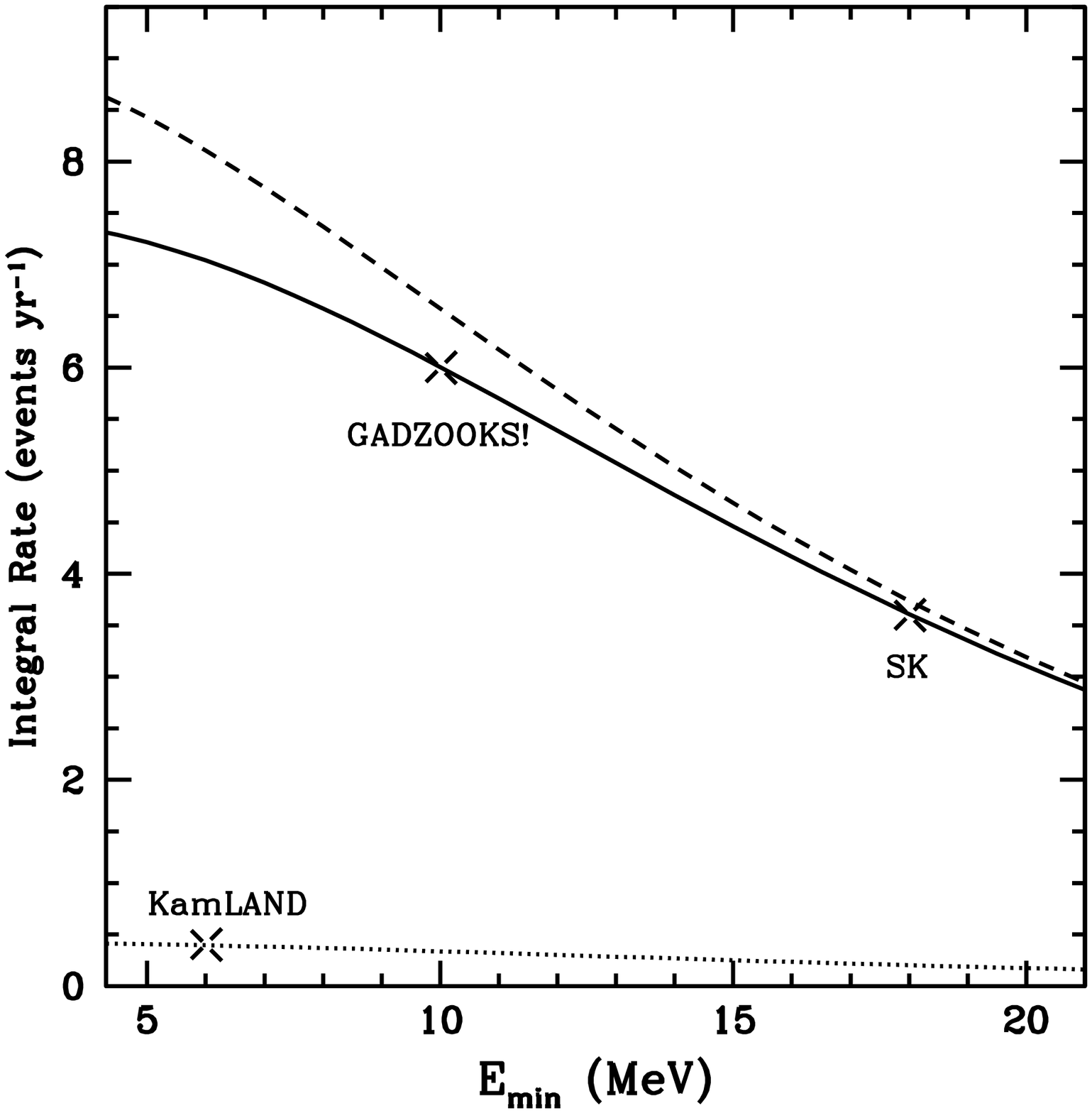,width=8cm}} 
\end{center} 
\caption{Left: Estimates of the supernova relic $\barnue$ integral 
  flux for positron  energies $E > E_{\rm min}$, as a function of 
  $E_{\rm min}$.  As in Figure 1, the solid curve is our ``median 
  SDSS" model with a normalization chosen to saturate the \SK~upper 
  bound.  The dotted curve, also saturating the \SK~upper bound, 
  shows the effect of increasing the slope of the $z > 1$ SN rate. 
  The \SK~upper bound is indicated by the cross. The dashed curve 
  is from an estimated lower bound to the SN rate based on the SDSS 
  results. The Kamland and \SK ~threshholds are shown by the vertical 
  lines. Right: Estimates of the SRN $\barnue$ integral event rates for 
  positron energies $E > E_{\rm min}$, as a function of $E_{\rm min}$.  
  The dashed, dotted, and upper solid curves correspond to those for \SK~in 
  Figure \ref{fig:SNrate}. The lower solid curve corresponds to the solid 
  curve in Figure \ref{fig:SNrate} for the integral event rate at KamLAND. 
  Crosses mark the levels of the predicted event rates for our best estimate 
  for each detector. 
\label{fig:flux}} 
\end{figure}

Having assembled the necessary ingredients, we are now in a position 
to calculate the SRN flux at Earth and to estimate the event rates at 
\kam~and \SK. For our standard cases we assume a normal mass hierarchy 
(no mixing inside the supernovae) and choose tan$^{2}\theta_{\odot} =  
0.46$ ($\sin^{2}(\theta_{\odot}) = 0.31$), in the middle of the solar  
neutrino and \kam~range (see \S \ref{sec:mixing}).  For \SK~we find  
\begin{equation}  
0.3 \, \lsim \, F(E > 18~{\rm MeV}) \, \lsim \, 1.2 ~\barnue~ {\rm cm}^{-2} \, 
{\rm s}^{-1}, 
\end{equation} 
while for \kam, 
\begin{equation} 
1.9 \, \lsim \, F(E > 6~{\rm MeV}) \, \lsim \, 7.8 ~\barnue~ {\rm cm}^{-2} \, 
{\rm s}^{-1}.  
\end{equation} 
  The upper and lower limits correspond to the solid and dashed curves 
in Figure \ref{fig:flux} respectively. For an inverted mass hierarchy, 
the detectable flux for $E > 18$ MeV increases by $\sim 50\%$, as this 
interval samples the high-energy tail of the SN neutrino spectrum  
(locally and, especially, at higher redshifts).  If this inverted 
hierarchy were realized, a lower supernova rate would be required in 
order to remain consistent with the \SK~upper limit.  An inverted 
hierarchy is less important for the interval $E > 6$ MeV, which is  
sensitive to more of the (zero-redshift) spectrum, and for our SN 
neutrino spectrum and SFR parameters there is a negligible change 
in the detectable rate.  Note that the median/best estimate SFR 
(solid curve in Figure \ref{fig:flux}) saturates the current \SK~upper 
bound of $1.2~\barnue~$cm$^{-2}$s$^{-1}$ \cite{SK}, suggesting that 
with a further reduction in the background, \SK~may detect the SRN 
background (see below for further discussion).   
 
In order to calculate event rates corresponding to these neutrino fluxes, 
the relevant cross sections and energy thresholds are required.  Neglecting 
the smearing of events due to finite energy resolution, the event rate for 
positrons produced by the SRN with neutrino energies between $E_1$ and 
$E_2$ (assuming 100\% efficiency) is:   
\begin{equation}   
{\rm R}(E_1,E_2)= {\rm N}_{{\rm p}} 
\int_{E_1+1.3 {\rm MeV}}^{E_2 + 1.3 {\rm MeV}}   
\sigma(E) \frac{d F}{d E} d E \,,  
\label{eq:eventrate}   
\end{equation}   
where $E$ is the $\barnue$ energy, ${\rm N}_{{\rm p}}$ is  
the number of protons in the fiducial mass of the detector and $\sigma(E)$ 
is the $\barnue$ cross section on protons \cite{CrossSection}. Here we take  
$E_2 = 82 \, {\rm MeV}$. In the left panel of Figure (\ref{fig:flux}) we 
show the results for the integral event rates for \kam, \SK, and SK-GADZOOKS. 
The results for \SK~cover the range of SN rates allowed by combining the 
SDSS and SK limits, while for \kam~we show the integral event rate for our 
best estimate.  We estimate a \SK~event rate   
\begin{equation}  
 1.1 \le {\rm Rate} \, (E_1 > 18 \, {\rm MeV}) \, \le 3.6 \, 
{\rm events}/{\rm yr},  
\label{eq:SKrate} 
\end{equation}  
and a \kam ~event rate 
\begin{equation}  
 0.1 \le {\rm Rate} \, (E_1 > 6 \, {\rm MeV}) \, \le 0.4 \, 
{\rm events}/{\rm yr}.   
\label{eq:KAMrate} 
\end{equation}   
 
\begin{figure}   
\begin{center}  
\epsfxsize=3.2in  
\epsfbox{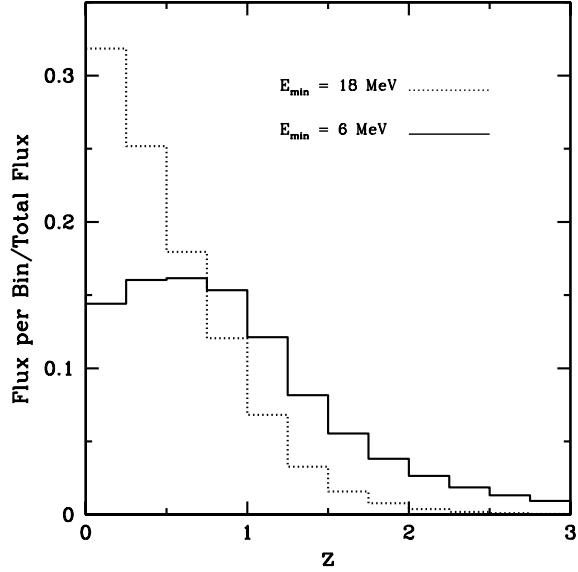} 
\end{center}   
\caption{The fraction of the relic neutrino flux contributed  
by \snii~in each redshift interval. The dotted histogram is  
for \SK, while the solid histogram is for KamLAND.}  
\label{fig:nufrac} 
\end{figure}

Although \kam~is a much smaller detector, it does have 
the advantage of no background and sensitivity to lower energies, 
permitting it to sample higher redshift supernovae and more of the 
energy spectrum of lower redshift supernovae.  This effect is 
illustrated in Figure~(\ref{fig:nufrac}) where the fractions of  the SRN 
fluxes contributed by \snii~at different redshifts are compared for 
\kam~($E_{1} = 6$~MeV) and \SK~($E_{1} = 18$~MeV). However, if the SFR has  
been underestimated at  
higher redshifts, KamLAND will detect even more relic neutrinos.   
For example, increasing the slope of the high-$z$ supernova rate from $\alpha  
= 0$ to 2 results in a $ \sim 36\%$ increase in flux for $E_1 > 6$~MeV,  
but only a $ \sim 4\%$ change in the event rate for the \SK~window   
$E_1 > 18$~MeV. 

What if both \kam~and \SK~ should detect the SRN?  This would open 
up the possibility of learning about the star formation history  
for $z ~\gsim 1$, which is currently not well understood.  A  
differential comparison of the \kam~and \SK~signals, the latter  
of which comes primarily from $z < 1$ (Fig.~(\ref{fig:nufrac})),  
would contain information about the $z > 1$ SRN rate and, therefore,  
about the $z > 1$ supernova rate.  It is informative to define  
the quantity 
\begin{equation} 
r = ({R_{\rm K}/V_{\rm K} \over R_{\rm SK/}V_{\rm SK}}) - 1,  
\label{eq:r} 
\end{equation} 
where $R_{K,SK}$ and $V_{K,SK}$ are the event rates and fiducial 
volumes at \kam~(K) and \SK. The parameter $r$ faithfully  
captures the effect of changes in $\rsn(z)$ on the ratio of the  
flux in the 6 -- 18 MeV window compared to that in the $>$ 18 MeV  
window.  Note that $r$ does not depend on the normalization $\rsn(0)$;  
it is only sensitive to the {\it shape} of the $\rsn$ -- redshift  
relation $\rsn(z)$.  Herein lies the sensitivity to supernovae  
at $z > 1$.  If the slope of the $z >1$ SFR is increased from  
$\alpha = 0$ to 2, $r$ changes by roughly 25\%.  For example, for 
the ranges of event rates discussed above, $1.8 < r <2.2$. 
However, changes  
in $\alpha$ are, to some extent, degenerate with changes in $\beta$.   
The advantages for studying the evolution of the cosmic SFR provided  
by the detection of the SRN at \kam~will come from combining such  
information with other, more traditional, astronomical observations  
of the cosmic SFR which can further restrict the range of $\beta$.   
Such analyses will provide unique information about the SFR at high  
redshift.  
 
\section{Prospects For Detection of the SRN} 
 
If the SRN background is indeed at the level of our median estimate 
(middle curve of Figure \ref{fig:flux}), how long will it take 
for \SK~and \kam~to detect the SRN?  To answer this question we must  
estimate the size of the error bars on the flux of SRN obtainable 
at \SK~and \kam, as a function of the number of years of data  
accumulated, $T$.  This expected error can then be compared to the  
likely ranges for the SRN-expected flux.  Our model for the error  
only includes the statistical errors, which in fact is a good approximation  
\cite{SK}.  The data, given the parameters which determine the SRN  
and the atmospheric neutrino fluxes, is expected to be Poisson  
distributed.  We neglect the correlations between bins due to energy  
resolution effects. If the bin sizes are larger than about twice the  
energy resolution, the correlations will change our results at  
the $\sim 10\%$ level.   
 
The recent \SK~ analysis \cite{SK} lists the detection efficiency for 
the SRN events as 47\% for  $E_1 \leq 34\,{\rm MeV}$ and 79\% for  
$E_1 > 34\,{\rm MeV}$ after full data reduction. We use an efficiency of 
unity for Kamland. We modify the event rate for the SRN and the 
backgrounds in accordance with this prescription for efficiency. The 
event rate in each energy bin is denoted by $S_{\alpha i}$ where   
$i$ refers to the energy bin. $\alpha$ identifies whether the 
event rate is due to the SRN, invisible muon decay, or 
the atmospheric neutrinos; $f_\alpha$ is the corresponding flux 
normalization.   
Assume that each bin accumulates $N_i$ events (after full data 
reduction) in $T$ years of runtime.  
The Poisson measurement error is $\sigma_i^{2} = N_i$. We define  
\begin{equation} 
F(\alpha,\beta) = T\, \Sigma_i \frac{T}{N_i} 
\frac{S_{\alpha i}}{f_\alpha}  
\frac{S_{\beta i}}{f_\beta} \,. 
\label{eq:fisher} 
\end{equation} 
The error on $f_\alpha$ is then given by 
$[F^{-1}(\alpha,\alpha)]^{1/2}$.  Assuming the model described above 
is a good fit to data (as demonstrated by the \SK~ results \cite{SK}), 
we forecast the errors on $f_\alpha$ by replacing $N_i$ by its 
expected mean value,  $T\Sigma_\alpha S_{\alpha i}$. 
 
With the above efficiency corrections, for the \SK~window $E_1 > 18$ MeV 
the expected 1-$\sigma$ error on the SRN event rate with $T$ years of data is  
$\sim 11 /\sqrt{T}$ events/yr. 
We note that with a total of $8-9$ years of data, \SK~will be able to make  
a 1-$\sigma$ detection if the SRN flux is close to our estimate from  
median SFR parameters (solid line in Figure \ref{fig:flux}). For \kam~, the  
primary background for the SRN is from atmospheric $\nu_{e}(\barnue)$, which  
has negligible expected rate in the range $\lsim \, 25$ MeV.  
To determine this background at \kam, we correct the results from \SK~  
\cite{Atmospheric} for the number of target protons in the 1 kton \kam~  
volume. We determine that \kam~can make a 1-$\sigma$ detection of the SRN  
flux with $\sim 5$ kton-years of data, if the flux is close to the solid  
curve in Figure \ref{fig:flux}.  
 
These estimated times to detection are long because the expected signal 
is close to the background when the detector is big and high threshold 
(\SK) or small and low background (\kam).   We note that if \SK~can 
sufficiently reduce the background from spallation and sub-Cerenkov 
$\mu^{\pm}$ decays, it can become an optimal SRN detector, opening 
up the $E > 10$ MeV window for detection of the SRN \cite{GADZOOKS}. 
Assuming $80\%$ efficiency for detecting the SRN for $E > 10$ MeV 
(M. Vagins, private communication), and a factor of 5 reduction in 
the backgrounds \cite{GADZOOKS}, we determine that \SK~ will make an 
immediate detection of the SRN background, at minimum a 1-$\sigma$ 
detection with less than one year of data. This opens up the possibility 
that we could constrain the space of fundamental \snii~ parameters 
which affect the spectrum of neutrino flux from a \snii~significantly. 
The general strategy would be to relax some assumption (motivated by 
simulations) about the neutrino spectrum and test them with SRN data. 
 
\section{Conclusions} 
 
Stimulated by  the \SK~collaboration's \cite{SK}  remarkable reduction 
in their backgrounds  for detection of the SN  relic neutrinos ($E \ge 
18$~MeV),  by  the  absence  of  significant backgrounds  to  the  SRN 
detection at \kam~\cite{kam2}(for  $E  >  6$~MeV),  and  by  recent 
progress in  pinning down the  cosmic star formation history  from new 
observations  including   the  Sloan  Digital  Sky   Survey,  we  have 
calculated the  expected fluxes and  event rates at \SK~ and KamLAND. 
The SRN flux presented here, based on the SN  rate derived from the star  
formation data, saturates the current \SK~  upper bound, suggesting 
that \SK~ may be  close to observing the SRN. For our median model 
(see  Fig. \ref{fig:flux}),  we conclude that the predicted flux at \SK~  
saturates the current experimental upper limit, with a corresponding event  
rate of $3.6 \, {\rm yr}^{-1}$. In this case \SK~ will  be  able to  make a   
1-$\sigma$ detection by doubling their data set, i.e. with a total of  
about 9 years (including the existing 4 years of data).   If  \SK~does  see   
the  SRN  background  flux, assuming that \kam~is $100\%$ efficient and can 
use the entire 1 kton volume for detection of the SRN, the associated event  
rate  expected at \kam~is small, $\sim 0.4 \, {\rm yr}^{-1}$. However, in this  
ideal scenario, the lack of any significant competing  backgrounds  
will permit  a 1-$\sigma$ detection of the SRN with only 5 years of data. 
 
\section*{Acknowlegments}   
 
G.S. thanks John Learned for encouragement and several discussions  
which helped to stimulate this investigation.  We thank John  
Beacom  and Mark Vagins for discussions about the supernova relic neutrinos and 
their GADZOOKS proposal.   
We thank Sinichiro Ando for correcting an equation in an earlier draft of  
this paper.  
The research of G.S., L.S., and T.P.W.  are supported at OSU by the 
Department of Energy grant DE-FG02-91ER40690;  M.K. is supported  
at U. C. Davis from the NSF and from NASA grant NAG5-11098 .

\end{document}